\documentclass[12pt]{article}
\usepackage{graphicx}
\usepackage{a4wide}
\usepackage{subfigure}
\usepackage{amsfonts,amssymb,amsmath}
\usepackage{mathptmx}

\newcommand{\eps}{\epsilon}

\newcommand{\cS}{{\mathcal S}}

\newcommand{\bdm}{\begin{displaymath}}
\newcommand{\edm}{\end{displaymath}}
\newcommand{\beq}{\begin{equation*}}
\newcommand{\eeq}{\end{equation*}}
\newcommand{\beqa}{\begin{eqnarray*}}
\newcommand{\eeqa}{\end{eqnarray*}}

\newcommand{\s}{^\prime}
\newcommand{\vr}{{r}}
\newcommand{\epse}{\epsilon}

\newcommand{\epsS}{\epsilon_{\rm s}}
\newcommand{\epsB}{\epsilon_{\rm B}}
\newcommand{\vOe}{{\Omega}}

\newcommand{\vO}{{\Omega}}

\newcommand{\rhoe}{\rho_{\rm in}}
\newcommand{\rhoc}{\rho_{\rm el}}

\newcommand{\Moller}{M{\o}ller }

\title{Deterministic Partial Differential Equation Model for Dose Calculation in Electron Radiotherapy}

\author{
Roland Duclous\footnote{CELIA \& IMB Laboratories, Bordeaux University, 33405 Talence, France, {\tt duclous@celia.u-bordeaux1.fr}} 
\and
Bruno Dubroca\footnote{CELIA \& IMB Laboratories, Bordeaux University, 33405 Talence, France, {\tt dubroca@math.u-bordeaux1.fr}} 
\and 
Martin Frank\footnote{RWTH Aachen University, Department of Mathematics \& Center for Computational Engineering Science
Schinkelstr.\ 2, 52062 Aachen, Germany, {\tt frank@mathcces.rwth-aachen.de}}
}

\begin{document}

\maketitle

\begin{abstract}
Treatment with high energy ionizing radiation is one of the main
methods in modern cancer therapy that is in clinical use. During the
last decades, two main approaches to dose calculation were used, Monte
Carlo simulations and semi-empirical models based on Fermi-Eyges
theory. A third way to dose calculation has only recently attracted
attention in the medical physics community. This approach is based on
the deterministic kinetic equations of radiative transfer. Starting
from these, we derive a macroscopic partial differential equation
model for electron transport in tissue. This model involves an angular
closure in the phase space. It is exact for the free-streaming and the
isotropic regime. We solve it numerically by a newly developed HLLC
scheme based on \cite{BerCharDub}, that exactly preserves key
properties of the analytical solution on the discrete level.  Several
numerical results for test cases from the medical physics literature
are presented.
\end{abstract}

\section{Introduction}
%
Mathematical methods play an increasing role in medicine, especially
in radiation therapy. Several special journal issues have been devoted
to cancer modeling and treatment, {\it
  cf.}\ \cite{BelMai05,BelMai06,BelMai07,Cen08} among others.

Together with surgery and chemotherapy, the use of ionizing radiation 
is one of the main tools in the therapy
of cancer. The aim of radiation treatment is to deposit enough energy
in cancer cells so that they are destroyed. On the other hand, healthy
tissue around the cancer cells should be harmed as little as
possible. Furthermore, some regions at risk, like the spinal chord,
should receive a dose below a certain threshold.

Most dose calculation algorithms in clinical use rely on the
Fermi--Eyges theory of radiation which is insufficient at
inhomogenities, {\it e.g.}\ the lung. This work,
on the other hand, starts with a Boltzmann transport model for the
radiation which accurately describes all physical interactions.

Until recently, dose calculation using a Boltzmann transport equation
has not attracted much attention in the medical physics
community. This access is based on deterministic transport equations
of radiative transfer. Similar to Monte Carlo simulations it relies on
a rigorous model of the physical interactions in human tissue that can
in principle be solved exactly. Monte Carlo simulations are widely
used, but it has been argued that a grid-based Boltzmann solution
should have the same computational complexity \cite{Bor98}.  Electron
and combined photon and electron radiation were studied in the context
of inverse therapy planning, cf.~\cite{TerKolVauHeiKai99,TerKol02} and
most recently \cite{TerVauBom08}. A consistent model of combined
photon and electron radiation was developed \cite{HenIzaSie06} that
includes the most important physical interactions. Furthermore,
several neutral particle codes have been applied to the dose
calculation problem, see \cite{GifHorWarFaiMou06} for a review and
most recently \cite{VasWarDav09}.

In this paper, we want to study a macroscopic approximation to the
mesoscopic transport equation. After the problem formulation in
section \ref{sec:DET}, we derive the approximation of the macroscopic
model in section \ref{sec:M1}. This approximation consists of a system
of nonlinear hyperbolic partial differential equations, whose
properties we briefly discuss. Due to the possibility of shock
solutions, hyperbolic PDEs have to be solved with great care. In
section \ref{sec:NumMeth}, we introduce a scheme which is adapted to
the problem at hand. Numerical results for tests from the medical
physics literature are presented in section \ref{sec:NumRes}.

\section{A deterministic model for dose calculation}
\label{sec:DET}
%
A ray of high energy electrons that interacts with human tissue is subject to elastic scattering processes and inelastic ones. It is this latter process that leads to energy deposition in the tissue {\it i.e.}~to absorbed dose. 

To formulate a transport equation for electrons we study their fluence
in phase space. Let $\psi(r,\eps,\Omega)\cos\Theta dA d\Omega d\epse
dt$ be the number of electrons at position $r$ -$r$ being a vector in
2D or 3D space- that move in time $dt$ through area $dA$ into the
element of solid angle $d\Omega$ around $\Omega$ with an energy in the
interval $(\epse, \epse+d\epse)$. The angle between direction $\vO$
and the outer normal of $dA$ is denoted by $\Theta$. The kinetic
energy $\epse$ of the electrons is measured in units of $m_ec^2$,
where $m_e$ is the electron mass and c is the speed of light.
\subsection{Boltzmann transport equation}
The transport equation can generally be formulated as \cite{Davison}
\begin{eqnarray}\label{electronEq}
\vO\cdot\nabla
\psi(r,\eps,\Omega)&=&
\rhoe(\vr)\int\limits_{\eps}^{\infty}\int\limits_{\cS^2}{\sigma}_{\rm in}(\epse\s,\epse,\vO\s\cdot\vO)\psi(r,\eps',\Omega')d\Omega^\prime d\epse\s\nonumber\\
&+&\rhoc(\vr)\int\limits_{\cS^2}\sigma_{\rm el}(\vr,\epse,\vO\s\cdot\vO)\psi(r,\eps,\Omega')d\Omega\s\nonumber\\
&-&\rhoe(\vr)\sigma_{\rm in}^{\rm tot}(\epse)\psi(r,\eps,\Omega)\nonumber\\
&-&\rhoc(\vr)\sigma_{\rm el}^{\rm tot}(\vr,\epse)\psi(r,\eps,\Omega),
\end{eqnarray}
with ${\sigma}_{\rm in}$ being the differential scattering cross section for inelastic scattering, and $\sigma_{\rm el}$ the differential cross section for elastic scattering; $\sigma_{\rm in}^{\rm tot}=\int_{S^2}{\sigma}_{\rm in}d\Omega $ and $\sigma_{\rm el}^{\rm tot}=\int_{S^2}{\sigma}_{\rm el}d\Omega$ are the total cross sections for inelastic and elastic scattering, respectively; $\rhoe$ and $\rhoc$ are the densities of the respective scattering centers. 

Explicit formulas for the cross sections that we used in this model can be found in section \ref{sec:CS}. They are based on the model developed in \cite{HenIzaSie06}. The energy integration is performed over $(\eps,\infty)$ since the electrons lose energy in every scattering event. Also, we consider only electron radiation. Equation (\ref{electronEq}) could also be used to model electrons which are generated by the interactions of photons with matter, as in \cite{HenIzaSie06}. In this case we would have an additional source term on the right hand side for the generated electrons.

Besides the transport equation one needs an equation for the absorbed dose. It was derived in \cite{HenIzaSie06} as an asymptotic limit of a model with a finite lower energy bound $\epsS > 0$. The formula is exact if one chooses the lower energy limit $\epsS=0$, as we do here.
\begin{equation} \label{dose}
D(\vr)=\frac{T}{\rho(\vr)} \int_{0}^{\infty} S(\vr,\epse\s)\psi^{(0)}(\vr,\epse\s)d\epse\s
\end{equation}
with 
\begin{equation}
\psi^{(0)}(\vr,\epse):=\int\limits_{S^2}\psi(\vr,\epse,\vO\s)d\Omega', \nonumber 
\end{equation}
$T$ being the duration of the irradiation of the patient and $\rho$ the mass density of the irradiated tissue. If all quantities are calculated in SI units, equation (\ref{dose}) leads to SI units J/kg or Gray (Gy) for the dose.

$S$ is the stopping power related to the inelastic cross section. It is defined as
\beq
S(\vr,\epse)=\rhoe(\vr)\int\limits_{0}^{\epse}\epse\s\sigma_{\rm in}(\epse,\epse\s) d\epse\s.
\eeq

\subsection{Continuous slowing-down approximation}
%
Electron transport in tissue has very distinctive properties. The soft
collision differential scattering cross sections have a pronounced
maximum for small scattering angles and small energy loss. This allows
for a simplification of the scattering terms in the Boltzmann
equation. The Fokker-Planck equation is the result of an asymptotic
analysis for both small energy loss and small deflections. It has been
rigorously derived in~\cite{Pom92} and has been applied to the above
Boltzmann model in~\cite{HenIzaSie06}.  However, some electrons will
also experience hard collisions with large changes in direction and
energy losses which have to be described by Boltzmann integral
terms. Thus we only use an asymptotic analysis to describe energy
loss, called continuous slowing-down approximation. This approximation
has a greater domain of validity than the Fokker-Planck
approximation. The Boltzmann equation in continuous slowing-down
approximation (BCSD) is \cite{LarMifFraBru97}
\begin{eqnarray}
\label{eq:BCSD}
\vO\cdot\nabla
\psi(r,\eps,\Omega)&=&
\rhoe(\vr) \int\limits_{\cS^2}{\sigma}_{\rm in}^{\rm CSD}
(\epse,\vO\s\cdot\vO)\psi(r,\eps,\Omega')d\Omega^\prime \nonumber\\
&+&\rhoc(\vr)\int\limits_{\cS^2}\sigma_{\rm el}
(\vr,\epse,\vO\s\cdot\vO)\psi(r,\eps,\Omega')d\Omega\s\nonumber\\
&-&\rhoe(\vr)\sigma_{\rm in, \rm tot}(\epse)\psi(r,\eps,\Omega)\nonumber\\
&-&\rhoc(\vr)\sigma_{\rm el,\rm tot}(\vr,\epse)\psi(r,\eps,\Omega) \nonumber\\
&+& \frac{\partial}{\partial\eps} (S(r,\eps) \psi(r,\eps,\Omega))
\end{eqnarray}
with 
$$
{\sigma}_{\rm in}^{\rm CSD} = \int_{0}^{\infty} \sigma_{\rm in}(\eps,\eps',\mu) d\eps'.
$$
A truncation in the energy space is introduced, that does not allow particles with arbitrary high energy,
\begin{equation}
	\lim_{\eps\to\infty}\psi(r,\epse,\Omega) = 0. 
\end{equation}
In the numerical simulations, we use a sufficiently large cutoff energy. 
Furthermore, we prescribe the ingoing radiation at the spatial boundary,
\begin{equation}	
	\psi(r,\eps,\Omega) = \psi_b(r,\eps,\Omega) \quad\textrm{for}\quad n\cdot\Omega<0,
\end{equation}
where $n$ is the unit outward normal vector.

\subsection{Modeling of Scattering Cross Sections}\label{sec:CS}
%
\subsubsection{Henyey-Greenstein Scattering Theory}
The detailed interactions of electrons with atoms give rise to complicated
explicit formulas for the scattering coefficients. Because of this, many studies use the
simplified Henyey-Greenstein scattering kernel for elastic scattering~\cite{AydOliGod02},
\begin{equation}\label{kernel}
    \sigma_{HG,g}(\mu) = \frac{1-g^2}{4\pi(1+g^2-2g\mu)^{3/2}}.
\end{equation}
The parameter $g$, which can depend on $r$, is the average cosine of
the scattering angle and is a measure for the anisotropy of the
scattering. The case where $g \leq 1$ matches an anisotropic scattering
configuration.

\subsubsection{Mott and \Moller Scattering}
A more realistic model for elastic and inelastic scattering of
electrons in tissue has been developed in \cite{HenIzaSie06}.  This
model introduces material parameters (namely densities $\rho_e$ and
$\rho_c$, ionization energy $\eps_B$ and effective atomic charge
$Z$). The energy integration for inelastic scattering is cut-off at
$\epse_B$.

The model uses the Mott scattering formula for elastic scattering of an
electron by an ion \cite{MottMassey,Lehmann} \beq \sigma_{\rm
  Mott}(\vr,\epse,\vOe\s\cdot\vOe)=\frac{Z^2(\vr)r_{\rm
    e}^2(1+\epse)^2}{4[\epse(\epse+2)]^2(1+2\eta(\vr,\epse)-\cos\vartheta)^2
} \left[1-\frac{\epse(\epse+2)}{(1+\epse)^2}\sin^2\frac{\vartheta}{2}
  \right], \eeq with $\vartheta = \arccos(\vOe\s\cdot\vOe)$, and
$\epse$ is the outcoming electron energy in $m_ec^2$ units.  Here,
$\alpha\approx 1/137$ is the fine structure constant, $Z$ is the
atomic number of the irradiated medium, $r_{\rm e}$ is the classical
electron radius. $Z$ depends on $\vr$ to account for heterogeneous
media. To avoid an otherwise occurring singularity at $\vartheta=0$ a
screening parameter
$$
\eta(r,\epse) = \frac{\pi^2 \alpha^2 Z^{2/3}(r)}{\eps(\eps+2)} ,
$$ can be introduced \cite{ZerKel67} that models the screening effect
of the electrons of the atomic shell, denoted by $a$.

The inelastic scattering process is \Moller scattering, where an
electron impinges an atom that releases itself an additional
electron
$$ e^- + a \rightarrow a^+ + 2 \ e^- $$ For this process, the
electrons can be considered indistinguishable. The electron which has
the higher energy after the collision is called primary electron, the
other electron secondary. Due to kinematical reasons of the scattering
processes the range of solid angles in \Moller scattering is
restricted. After the collision, the angle between the directions of
the electrons is at most $\pi/2$. For an angle in $[0,\pi/4]$, the
electron with energy $\eps$ is the primary electron, for an angle in
$[\pi/4,\pi/2]$, it is the secondary electron. Therefore the \Moller
cross section can be written as \beqa \sigma_M = \tilde{\sigma}_M
\chi_{\{ 0<\Omega\cdot\Omega'<\sqrt{2}/2\}} +
\tilde{\sigma}_{M,\delta} \chi_{\{\sqrt{2}/2<\Omega\cdot\Omega'<1\}},
\eeqa where $\chi$ denotes the characteristic function of a set,
$$
\tilde{\sigma}_{\rm M}(\epse\s,\epse,\vOe\s\cdot\vOe)=\sigma_{\rm M}(\epse\s,\epse)\delta_{M}(\mu,\mu_p)\frac{1}{2\pi},\quad \mu=\vOe\s\cdot\vOe ,
$$ 
is the \Moller differential cross section of primary electrons and 
$$
\tilde{\sigma}_{\rm M,\delta}(\epse\s,\epse,\vOe\s\cdot\vOe)=\sigma_{\rm M}(\epse\s,\epse)\delta_{\rm M,\delta}(\mu,\mu_\delta)\frac{1}{2\pi},\quad \mu=\vOe\s\cdot\vOe ,
$$ 
is the \Moller differential cross section of secondary electrons. Here,
$$
\sigma_{\rm M}(\epse\s,\epse)=\frac{2\pi r_{\rm e}^2(\epse\s+1)^2}{\epse\s(\epse\s+2)}
           \left[  \frac{1}{\epse^2}+\frac{1}{(\epse\s-\epse)^2}
                  +\frac{1}{(\epse\s+1)^2}-\frac{2\epse\s+1}{(\epse\s+1)^2\epse(\epse\s-\epse)}  \right] ,
$$
and
\begin{align*}
\delta_{M}(\mu_{\rm e},\mu_p) &= \delta\left(\mu_{\rm e}-\sqrt{\frac{\epse}{\epse\s}\frac{\epse\s+2}{\epse+2}}\right),\text{ for }\epse>\frac{(\epse\s-\epsB)}{2} , \\
\delta_{\rm M,\delta}(\mu_{\rm e},\mu_\delta) &=\delta\left(\mu_{\rm e}-\sqrt{\frac{\epse}{\epse\s}\frac{\epse\s+2}{\epse+2}}\right),\quad\epse<\frac{(\epse\s-\epsB)}{2}.
\end{align*}
In the simulations the model parameters $\rho_{\rm el}$, $\rho_{\rm
  in}$, $\epse_B$ and $Z$ are fitted to tabulated values taken from
the database of the PENELOPE Monte Carlo code \cite{SalFerSem08}.

\section{Partial Differential Equation Model}
\label{sec:M1}
%
We will try to reduce the cost of solving system (\ref{electronEq}) by assuming a minimum
entropy principle for the angle distribution of particles. This principle has been first proposed by Jaynes
\cite{Jay57} as a method to select the most likely state of a thermodynamical system having only incomplete information. It has subsequently been developed in \cite{Min78}, \cite{Lev84}, \cite{AniPenSam91} and \cite{DubFeu99}, among others, and has become the main concept of rational extended thermodynamics \cite{MullerRuggeri}. A full account and an exhaustive list of references on the historical development can be found in \cite{HauLevTit08}.

We define the first three moments in angle:
\begin{align}
\label{eq_def_mom}
\psi^{(0)}(r,\eps) &= \int_{S^2} \psi(r,\eps,\Omega)d\Omega , \\
\psi^{(1)}(r,\eps) &= \int_{S^2} \Omega\psi(r,\eps,\Omega)d\Omega , \\
\psi^{(2)}(r,\eps) &= \int_{S^2} (\Omega\otimes\Omega)\psi(r,\eps,\Omega)d\Omega,
\end{align}
where we note that $\psi^{(0)}$ is a scalar, $\psi^{(1)}$ is a vector and $\psi^{(2)}$ is a tensor.

\begin{figure}
  \subfigure[Eddington factor.]{
  \includegraphics[width=0.48\linewidth]{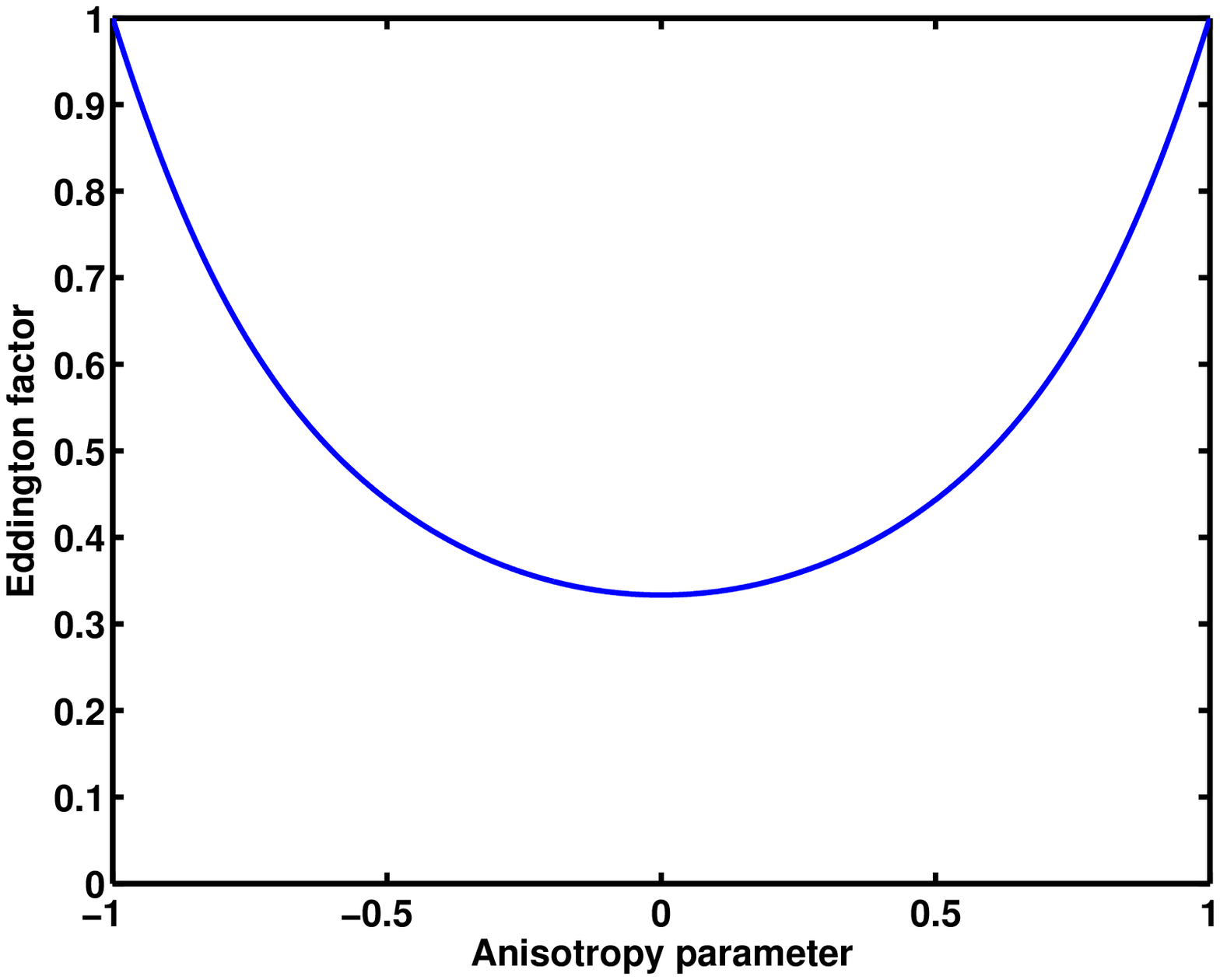}}
  \subfigure[Eigenvalues.]{
  \includegraphics[width=0.48\linewidth]{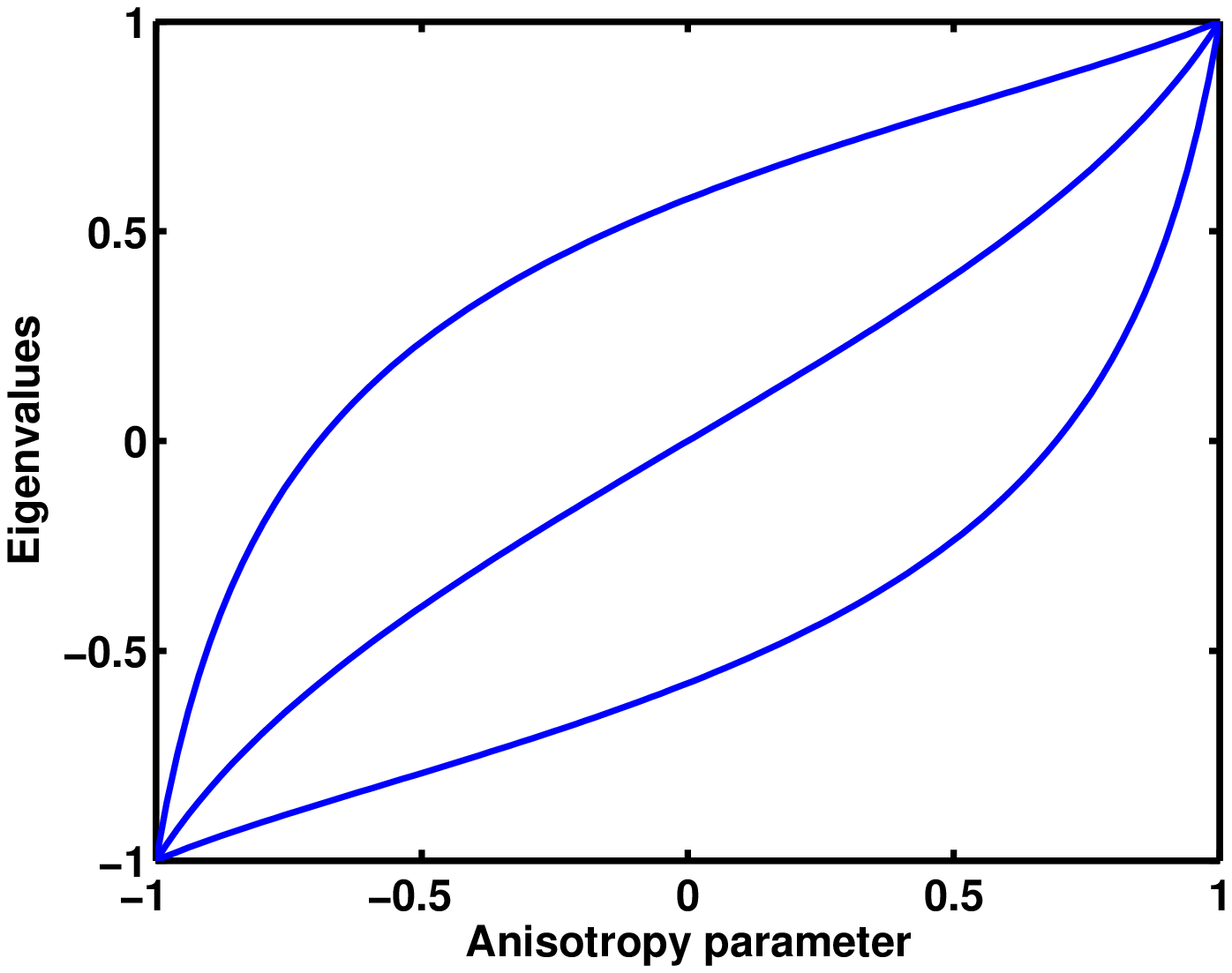}}
  \caption{Eddington factor $\chi$ and system eigenvalues versus anisotropy
    parameter $| \alpha |$.}
  \label{fig_def_chi}
\end{figure}

If we integrate the system (\ref{eq:BCSD}) over $\Omega$, we can derive the following
equations,
\begin{subequations}  \label{es_sys_mom1}
\begin{align}
\nabla_x \psi^{(1)} &= \frac{\partial}{\partial\eps}(S\psi^{(0)}) , \\ \label{es_sys_mom2}
\nabla_x \psi^{(2)} &= -(T_M+T_{\rm Mott})\psi^{(1)} + \frac{\partial}{\partial\eps}(S \psi^{(1)}).
\end{align}
\end{subequations}
We have introduced the transport coefficients
\begin{align}
	T_{\rm in}(r,\eps) &= \pi \rho_{\rm in}(r) \int_{\eps_B}^{(\eps-\eps_B)/2}\int_{-1}^1 (1-\mu)\sigma_{\rm in}(\eps,\eps',\mu)d\mu d\eps' , \\
	T_{\rm{el}}(r,\eps) &= \pi \rho_{\rm el}(r) \int_{-1}^1 (1-\mu)\sigma_{\rm el}(\eps,\mu)d\mu.
\end{align}
These coefficients and the stopping power can be computed for both Henyey-Greenstein and Mott/\Moller scattering. Explicit expressions can be found in \cite{HenIzaSie06,FraHenKla06}.

The remaining problem is the computation of moment $\psi^{(2)}$ as
a function of $\psi^{(0)}$ and $\psi^{(1)}$. The Minimum Entropy $M_1$
closure for electrons \cite{BruHol01} can be derived in the following
way. To close the system we determine a distribution function
$\psi_{ME}$ that minimizes the entropy of the electrons,
\begin{equation}
  H_R^\ast (\psi)= - \int_{S^2}  \psi \log \psi d\Omega ,
\end{equation}
under the constraint that it reproduces the lower order moments,
\begin{equation}
  \int_{S^2} \psi_{ME}d\Omega = \psi^{(0)} \quad\textrm{and}\quad \int_{S^2} \Omega\psi_{ME}d\Omega = \psi^{(1)}.
\end{equation}
By using this entropy, we have implicitly assumed that the electrons obey classical Maxwell-Boltzmann statistics. This is justified, since here quantum effects can be neglected.

Analogous to the calculations in \cite{Lev84} we can show that the entropy minimizer has the following
form,
\begin{equation}
  \label{eq_def_closure}
  \psi_{ME} = a_0 \exp(-\Omega \cdot a_1) ,
\end{equation}
where $a_0$ is a non-negative scalar, and $a_1$ is
a three component real valued vector. This is a Maxwell-Boltzmann type distribution and $a_0$, $a_1$ are (scaled) Lagrange multipliers enforcing the constraints. 
An important parameter is the anisotropy parameter $\alpha$,
$$
\alpha =\frac{\psi^{(1)}}{\psi^{(0)}} ,
$$
whose norm is by construction less than or equal to one. If we compute the different moments of the distribution function given by \eqref{eq_def_closure} we obtain,
\begin{eqnarray}
  \label{eq_calcul_f02}
  \psi^{(0)} = 4 \pi a_0 \frac{\sinh(|a_1|)}{|a_1|}, \ 
  \psi^{(1)} = 4 \pi a_0 \frac{\sinh(|a_1|) (1 - |a_1| \coth(|a_1|))}{|a_1|^3} \ a_1.
\end{eqnarray}
In fact, these relations can be combined to give,
\begin{equation}
  \label{eq_def_alpha1}
  \alpha = \frac{1 - |a_1| \coth(|a_1|)}{|a_1|^2} \ a_1,
\end{equation}
or by taking the modulus,
\begin{equation}
  \label{eq_def_alpha}
  |\alpha| = \frac{|a_1| \coth(|a_1|) - 1}{|a_1|}.
\end{equation}

The relation \eqref{eq_def_alpha} cannot be inverted explicitly by
hand, {\it i.e.}\ we cannot express $|a_1|$ as a function of $\alpha$ in a closed form. However, this relation determines a unique solution which can in principle be computed. If we
assume that we know $a_1$, $\psi^{(2)}$ can be computed as
\begin{equation}
  \label{eq_def_f2}
  \psi^{(2)} = \psi^{(0)} \left (\frac{1-\chi(\alpha)}{2} I
  + \frac{3 \chi(\alpha) - 1}{2} \alpha \otimes
  \alpha \right ),
\end{equation}
where
\begin{equation}
  \chi = \frac{|a_1|^2 - 2 |a_1| \coth(|a_1|) + 2}{|a_1|^2}
\end{equation}  
is a function of $\alpha$ by means of \eqref{eq_def_alpha}.

For its efficient numerical evaluation, the Eddington factor has to be approximated. Several possibilities exist: 
\begin{itemize}
\item One could solve the closure relation \eqref{eq_def_alpha} for $|a_1|$ {\it e.g.}\ by a Newton iteration in each step during the simulation. 
\item One could precompute a table that gives the Eddington factor $\chi$ as a function of $\alpha$. 
\item One could approximate $\chi(\alpha)$ by a suitable special function.
\end{itemize}
The second approach has been followed in \cite{FraHenKla06}. It is advantageous only if the space in which one interpolates is low-dimensional. For more moments, this approach becomes more expensive, and the first approach appears to be more advantageous. 

In some cases, an {\it ansatz} for $\chi$ can provide a good approximation. This is the approach we are following here.
The Eddington factor $\chi$ can be approximated by a very simple rational function,
\begin{equation}
  \label{defchi}
  \chi(\alpha) \approx  
  {\frac{  
      a_6\,{\alpha}^{6} +
      a_4\,{\alpha}^{4} + 
      a_2\,{\alpha}^{2} + 
      a_0
    }{
      {\alpha}^{4}      + 
      b_2\,{\alpha}^{2} + 
      b_0.
    }}
\end{equation}
This approximation is very accurate (the difference with exact curve is about
$10^{-15}$). The coefficients are given by
\begin{equation}
  \begin{array}{cc}
    a_0 = \ \ 0.762066949972264, & b_0 = \ \ 2.28620084991677,
    \\
    a_2 = \ \ 0.219172080193380, & b_2 = -2.10758208969840,
    \\
    a_4 = -0.259725400168378, &
    \\
    a_6 = \ \ 0.457105130221120. 
  \end{array}
\end{equation}

\section{Properties of the System}
\label{sec:PROP}
%
In the literature, the system that has been thoroughly investigated (both analytically and numerically) is system \eqref{es_sys_mom1} restricted to its conservative terms,  without external sources, but with time-dependence.

In the present work, we adapt a pseudo-time technique. We focus on the spatial discretization and use a standard discretization for the terms on the right-hand side. Thus we consider
\begin{subequations}  \label{time_sys_mom1}
\begin{align}
\frac{\partial}{\partial t} \psi^{(0)} + \nabla_x \psi^{(1)} &= 0,\\
\frac{\partial}{\partial t} \psi^{(1)} + \nabla_x \psi^{(2)}\left(\psi^{(0)},\psi^{(1)}\right) &= 0,
\end{align}
\end{subequations}
with the closure \eqref{eq_def_alpha}.

The Eddington factor $\chi$ is shown in Figure \ref{fig_def_chi}. Furthermore, we show the system eigenvalues in two dimensions. 
In the isotropic regime (anisotropy parameter zero), they coincide with the P1 eigenvalues. On the other hand,  in the case of free-streaming ($|\alpha|=1$), they coincide and have absolute value one. Thus the system (\ref{es_sys_mom1}) is hyperbolic and the speed of propagation is limited by one. Moreover the system is hyperbolic symmetrisable \cite{DubFeu99}.

System \eqref{time_sys_mom1} closed by the relation (\ref{eq_def_f2}) has been analyzed thoroughly in \cite{CouGou06}. There, solutions to Riemann problems are constructed and invariant regions are computed. Since the reconstruction (\ref{eq_def_closure}) of the kinetic distribution $\psi$ is always positive, it can be expected that system \eqref{eq_def_f2} - \eqref{time_sys_mom1} must admit a positive solution 
$\psi^{(0)}$ and a limited flux $\| \alpha \|<1$. To our knowledge, however, there exists no proof of this fact. The invariant regions computed in \cite{CouGou06} only cover a subset of all admissible values. For a related model \cite{FraPin05}, bounds were proved, but only in 1D and steady state. Nevertheless, we construct a scheme which preserves exactly the positivity of $\psi^{(0)}$ and the flux limitation, {\it i.e.}\ 
the convex set of the admissible states of the system \eqref{time_sys_mom1} is \cite{BerCharDub}
 $$
 \mathcal{A} = \left \{ \left ( \psi^{(0)},\psi^{(1)} \right ):\  \psi^{(0)} \geq 0,\ |\psi^{(1)}| \leq \psi^{(0)} \right \}.
 $$

In the absence of sources or boundaries, the total mass, momentum and energy are conserved.

In addition, the minimum entropy system recovers the equilibrium diffusion regime as a relaxation limit for large absorption coefficients \cite{CouGolGou05}.

In a two- or three-dimensional geometry, we have in addition \cite{BerCharDub}:
Let $n$ be the unit normal vector to
 an interface; then the system exhibits two acoustic waves, with
 velocities $\lambda_L(n)$ and $\lambda_R(n)$, supplemented by a
 contact wave with velocity $\beta(n)$. The quantity $\beta \cdot n$
 satisfies the following inequality $\lambda_L (n) \le \beta(n) \cdot
 n \le \lambda_R (n)$. The Riemann invariants associated with the
 contact wave are $\left \{ \beta , \Pi \right \} $. They are defined
 by the relations
\begin{subequations}  \label{riemann_inv}
\begin{align}
\psi_1 &= \left ( \Pi + \psi_0\right ) \beta ,\\
\psi_{2} &= \left ( \psi_0 + \Pi   \right )\beta \otimes \beta + \Pi Id \ .
\end{align}
\end{subequations}

\section{Numerical Method}
\label{sec:NumMeth}
%
The properties of the continuous model should be reproduced by the
numerical scheme. In particular the positivity and flux limitation
constraints are fundamental. An HLL scheme \cite{HartenLaxvanLeer} can be
constructed \cite{Batten,Buet,BerCharDub}, that satisfies the
required properties. However such an approach cannot capture the
contact discontinuity. To prevent this failure, an HLLC scheme
\cite{Batten} has been derived, that resolves the contact discontinuity and
satisfies the physical constraints.

To complete this presentation of
the numerical approximation, we mention that a suitable high order
extension that preserves both the positivity and the flux
limitation can be derived, relying on an appropriate limitation
procedure.

\subsection{An HLL scheme for the free transport $M_1$ angular moment system}
%
In this section, we derive a Finite Volume method, issued from the HLL
method \cite{HartenLaxvanLeer} to discretize the free transport equation 
\eqref{time_sys_mom1}. Put in other
words, we omit the source terms and we consider the one dimensional
generic conservative system
 \begin{eqnarray} 
  \label{following_form}
\displaystyle \frac{\partial}{\partial t}\mathcal{U}+ \frac{\partial}{\partial x} \left [ \mathcal{F} \left ( \mathcal{U} \right
)\right
] = 0 \ ,
\end{eqnarray} 
where 
$$
\mathcal{U}= \left ( \begin{array}{cc} \psi_0
  \\ \psi_1 \end{array} \right )
  $$ 
  and $\mathcal{F}$ stands for the
flux of the $M_1$ system in the $x$ space direction.\\ We consider a
structured mesh of size $\Delta x_i$, defined by the cells $I_i =
\left [x_{i-1/2} , x_{i+1/2} \right )$, where we have set
  $x_{i+1/2}=x_i+ \Delta x /2 \ , i \in \mathbb{Z} $, at time
  $t^n$. As usual, we consider known a piecewise contant approximation
  $U^h(x,t^n)$, defined by $U^h(x,t^n)=U_i^n, x \in I_i \ , \forall i
  \in \mathbb{Z}$.
  
\begin{figure}
  \begin{centering}
  \includegraphics[width=0.5\linewidth,height=4cm]{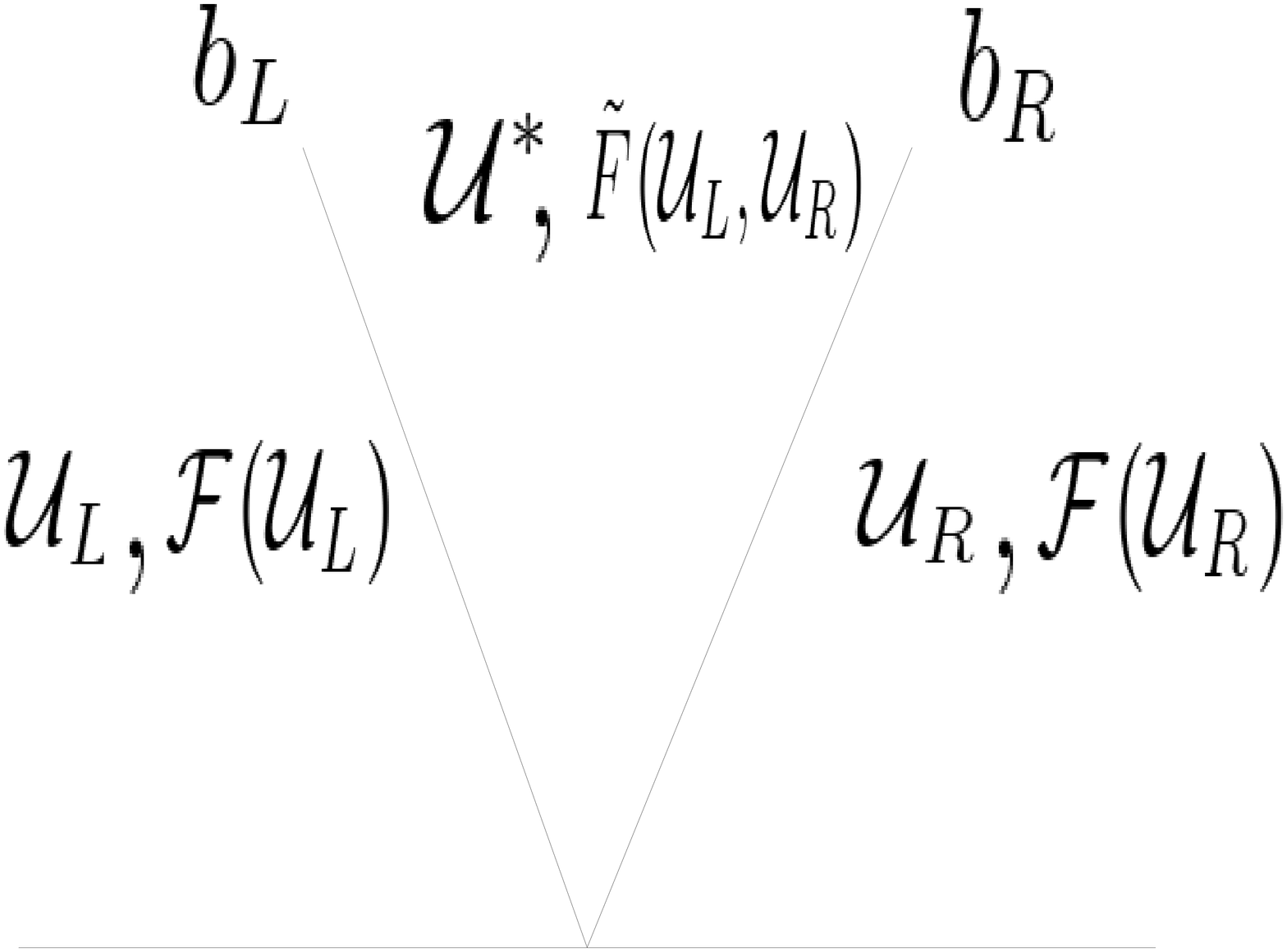}
  \caption{Structure of the approximate HLL Riemann Solver}
  \label{Riemann_fan}
 \end{centering}
\end{figure}

   At initial time $t=0$, we impose 
   $$
   U_i^0 =\displaystyle \int_{x_{i-1/2}}^{x_{i+1/2}}U_0(x)dx,
   $$ 
   where $U_0$ is the initial data. This approximation evolves in time,
  involving a suitable approximate Riemann solver. In the HLL
  approach, the exact Riemann solver solution is substituted by a
  single approximate state (see Figure \ref{Riemann_fan}).
Here $b_L$ and $b_R$ are relevant approximations of $\lambda_L$ and
 $\lambda_R$, respectively. Let us introduce the proposed approximate
 solution:
\begin{eqnarray} 
  \label{HLL_single_state}
{\mathcal{U}_{HLL}}(x,t) \equiv \left ( \begin{array}{cc} \psi_0(x,t)
  \\ \psi_1(x,t) \end{array} \right )_{HLL} = \left \{
 \begin{array}{ccc}  & {\mathcal{U}_L} &  \ \mbox{if} \ \frac{x}{t} < b_L   \ , \\
  & {\mathcal{U}^*} & \qquad \ \ \mbox{if} \ b_L \le\frac{x}{t} \le
   b_R \ , \\ & {\mathcal{U}_R} & \ \mbox{if} \ b_R <\frac{x}{t}
   \ .  \end{array} \right .
\end{eqnarray}
Moreover, the search of weak solutions leads to the Rankine-Hugoniot
jump conditions
\begin{subequations}\label{RH-jump}
\begin{align}   
    -  b_L \left [ \mathcal{U}^*- \mathcal{U}_L \right ]  +  \left [ \tilde{F}- \mathcal{F}(\mathcal{U}_L) \right ] &= 0, \\
  - b_R \left [ \mathcal{U}_R- \mathcal{U}^* \right ] +  \left [  \mathcal{F}(\mathcal{U}_R) - \tilde{F} \right ] & = 0.
\end{align}
\end{subequations}
These relations provide us with an explicit expression for the
intermediate state and flux of the numerical scheme
\begin{subequations}
\begin{eqnarray} 
  \label{app_state_flux}
\mathcal{U}^* & = & \frac{b_R \mathcal{U}_R-b_L \mathcal{U}_L-(\mathcal{F}(\mathcal{U}_R) -\mathcal{F}(\mathcal{U}_L)) }{b_R -b_L } \ ,
\\ \tilde{F}(\mathcal{U}_L,\mathcal{U}_R) & = & \frac{b_R 
  \mathcal{F}(\mathcal{U}_L) - b_L 
  \mathcal{F}(\mathcal{U}_R) - b_L  b_R 
  (\mathcal{U}_R-\mathcal{U}_L) }{b_R -b_L  } \ ,
\end{eqnarray}
\end{subequations}
At each interface $x_{i+1/2}$, we impose the above HLL approximate
Riemann solver, assuming the CFL like condition \eqref{CFLcond}
ensuring that the Riemann solvers do not interact in the case where $ b_{L,i+1/2} < 0 $ and $ b_{R,i-1/2}>0$ :
\begin{eqnarray} 
  \label{CFLcond}
\frac{\Delta t}{\Delta x} \le \frac{  b_{L,i+1/2}  b_{R,i-1/2}}{  b_{L,i+1/2}-  b_{R,i-1/2} } \ .
\end{eqnarray} We set
$\mathcal{U}^h(x,t+\Delta t)$, at time $t^n+ \Delta t$, the
superposition of the non-interacting Riemann solutions. We define the
updated approximation at time $t^{n+1}$ by 
$$ 
\displaystyle
\mathcal{U}_i^{n+1}= \frac{1}{\Delta x }
\int_{x_{i-1/2}}^{x_{i+1/2}}\mathcal{U}^h(x,t^n+\Delta t).
$$ 
An easy computation gives
\begin{eqnarray} 
  \label{genform}
\mathcal{U}_i^{n+1} =\mathcal{U}_i^{n} - \frac{\Delta t}{\Delta x}
({F}_{i+1/2}^n-{F}_{i-1/2}^n) \ ,
\end{eqnarray}
where 
$$
\displaystyle {F}_{i+1/2}^n(\mathcal{U}_i^n, \mathcal{U}_{i+1}^n)    = \left \{
 \begin{array}{ccc}  & \mathcal{F} \left (\mathcal{U}_i^n  \right ) &  \ \mbox{if} \ 0 < b_{L,i+1/2}    \\
  & \tilde{F}_{i+1/2}\left ( \mathcal{U}_{i}^n,\mathcal{U}_{i+1}^n
   \right ) & \qquad \qquad \ \ \mbox{if} \ b_{L,i+1/2} \le 0 \le
   b_{R,i+1/2} \\ & \mathcal{F} \left ( \mathcal{U}_{i+1}^n \right ) &
   \ \mbox{if} \ b_{R,i+1/2} < 0 \end{array} \right .  
   $$
  The
 robustness of the scheme, namely the positivity, the flux limitation,
 the total mass preservation, has been established for the HLL scheme (see
 \cite{BerCharDub} for further details).\\ Finally, concerning the
 high order extension, we adopt a van Leer MUSCL technique
 \cite{VanLeer}, supplemented by a suitable slope limitation
 preserving these expected physical properties \cite{Berthon_stab}.

\subsection{An accurate HLLC scheme}
%
 The HLL scheme has proved to be robust, however, its 2D extension
 fails when approximating contact waves. Several works
 \cite{Batten,BerCharDub} introduce a more accurate scheme, the HLLC
 scheme, based on a two state approximation, denoted by $U_L^*$ and
 $U_R^*$.
 
 First, let us recall the relevant linearization that
 permits us to define an approximation with two intermediate states: on
 the one hand, the Rankine-Hugoniot conditions \eqref{RH-jump} are
 considered; on the other hand, they are supplemented by the
 continuity of the Riemann invariants accross the contact wave:
\begin{eqnarray} 
  \label{cont_rie_inv}
\left ( \beta_x \right )_L^* = \left ( \beta_x \right )_R^* =
\beta_x^* \quad , \quad \Pi_L^*= \Pi_R^*= \Pi^* \ \ ,
\end{eqnarray}
where $ \beta_x$ and $\Pi$ are defined by the relation
\eqref{riemann_inv}. The combination of both the Rankine-Hugoniot
condition \eqref{RH-jump} and the relation \eqref{cont_rie_inv}
standing as the continuity of the Riemann invariants accross the
contact wave, is sufficient to determine uniquely
\cite{Batten,BerCharDub} the two approximate states $U_L^*$ and
$U_R^*$, together with their associated fluxes $\tilde{F}_L$ and
$\tilde{F}_R$. The proposed HLLC approximate solution can be written as
\begin{eqnarray} 
  \label{HLLC_two_state}
{\mathcal{U}_{HLLC}}(x,t) \equiv \left ( \begin{array}{cc} \psi_0(x,t)
  \\ \psi_1(x,t) \end{array} \right )_{HLLC} = \left \{
 \begin{array}{ccc}  & {\mathcal{U}_L} &  \ \mbox{if} \ \frac{x}{t} < b_L   \ , \\
  & {\mathcal{U}^*_L} & \qquad \ \ \mbox{if} \ b_L \le\frac{x}{t} \le
   \beta_x^* \ , \\ & {\mathcal{U}^*_R} & \qquad \ \ \mbox{if}
   \ \beta_x^* \le\frac{x}{t} \le b_R \ , \\ & {\mathcal{U}_R} &
   \ \mbox{if} \ b_R <\frac{x}{t} \ .  \end{array} \right .
\end{eqnarray}
Similar to the derivation of the HLL scheme, we integrate over a cell
$I_i$ the juxtaposition of the non-interacting HLLC Riemann
approximate solvers at each interface (projection step), in order to
obtain the updated quantity 
$$
\displaystyle \mathcal{U}_i^{n+1}=
\frac{1}{\Delta x }
\int_{x_{i-1/2}}^{x_{i+1/2}}\mathcal{U}^h(x,t^n+\Delta t).
$$  
This
brief description of the HLLC scheme is now completed. It is able to
capture exactly the contact wave, and satisfies the positivity, the
flux limitation, and the total mass preservation.

\section{Numerical Results}
\label{sec:NumRes}
%

\subsection{Central Void}
The first test case is taken from the medical physics literature
\cite{AydOliGod02}.  We consider only elastic scattering, which is
modeled by the Henyey-Greenstein kernel. Thus $S=0$ and $T_{\rm
  in}=0$.  We compare the particle flux $\psi^{(0)}(x)$ obtained with
the minimum entropy model (labeled M1) with a discrete ordinates
solution of the transport equation (labeled SN) with sufficiently many
angles (128). The method has been described in \cite{DubFra09}.

\begin{figure}
\centering  
\subfigure[Particle flux.]{\includegraphics[width=0.75\linewidth]{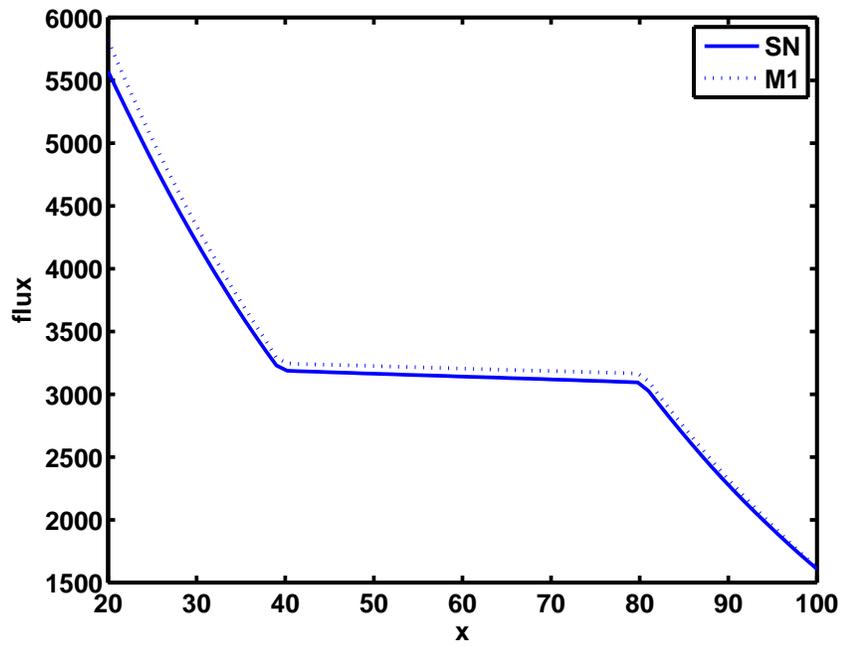}}
\subfigure[Comparison of $M_1$ and benchmark distribution functions. Logarithmic scale.]
{\includegraphics[width=0.75\linewidth]{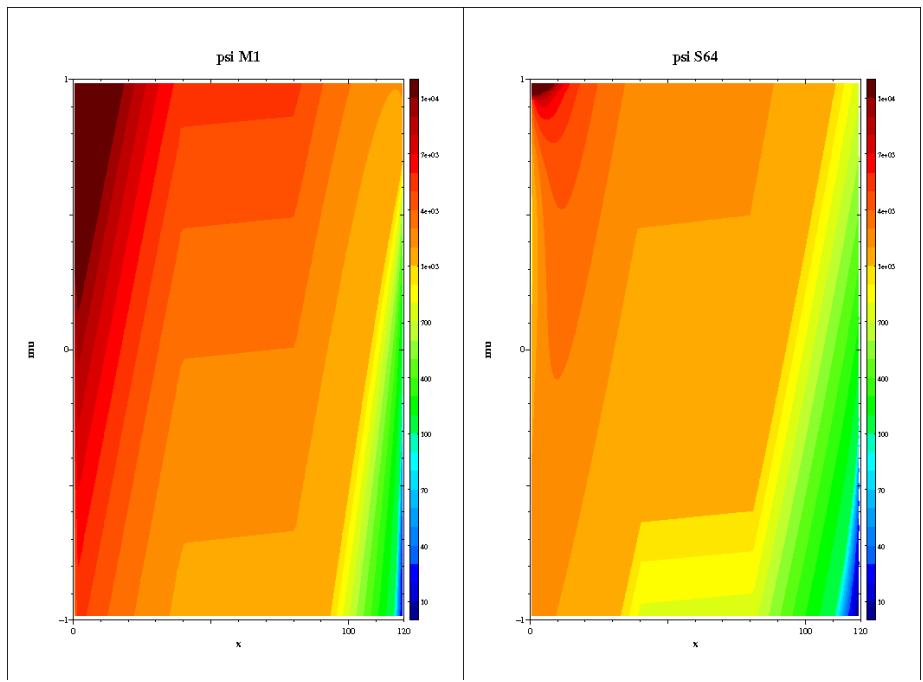}}
\caption{Geometry with central void.}\label{fig:WAW}
\end{figure}

The test case consists of a one-dimensional geometry with three
layers: optically thick, followed by optically thin followed again by
optically thick. The layers have an equal depth of 40 mm. The
scattering and absorption coefficients are $\sigma_s = 0.5$ mm$^{-1}$,
$\sigma_a=0.005$ mm$^{-1}$ for the optically thick region, and
$\sigma_s = 0.01$ mm$^{-1}$, $\sigma_a=0.0001$ mm$^{-1}$ for the
optically thin region. Moreover, $g=0$. Figure \ref{fig:WAW} shows the
particle flux $\psi^{(0)}$ as a functon of space. Compared to the
benchmark solution, the minimum entropy model slightly overestimates
but nevertheless quite accurately describes the particle flux. In
Figure \ref{fig:WAW} we also show the partice distribution function
$\psi(x,\Omega)$, where $\Omega = (0,0,\mu)$ in 1D. The main
difference is that for $M_1$, the forward-peak of the incoming
particles reaches further into the medium.

\subsection{Two-dimensional Void-like Layer}
\begin{figure}
  \subfigure[Geometry containing void layer.]{
  \includegraphics[width=0.38\linewidth]{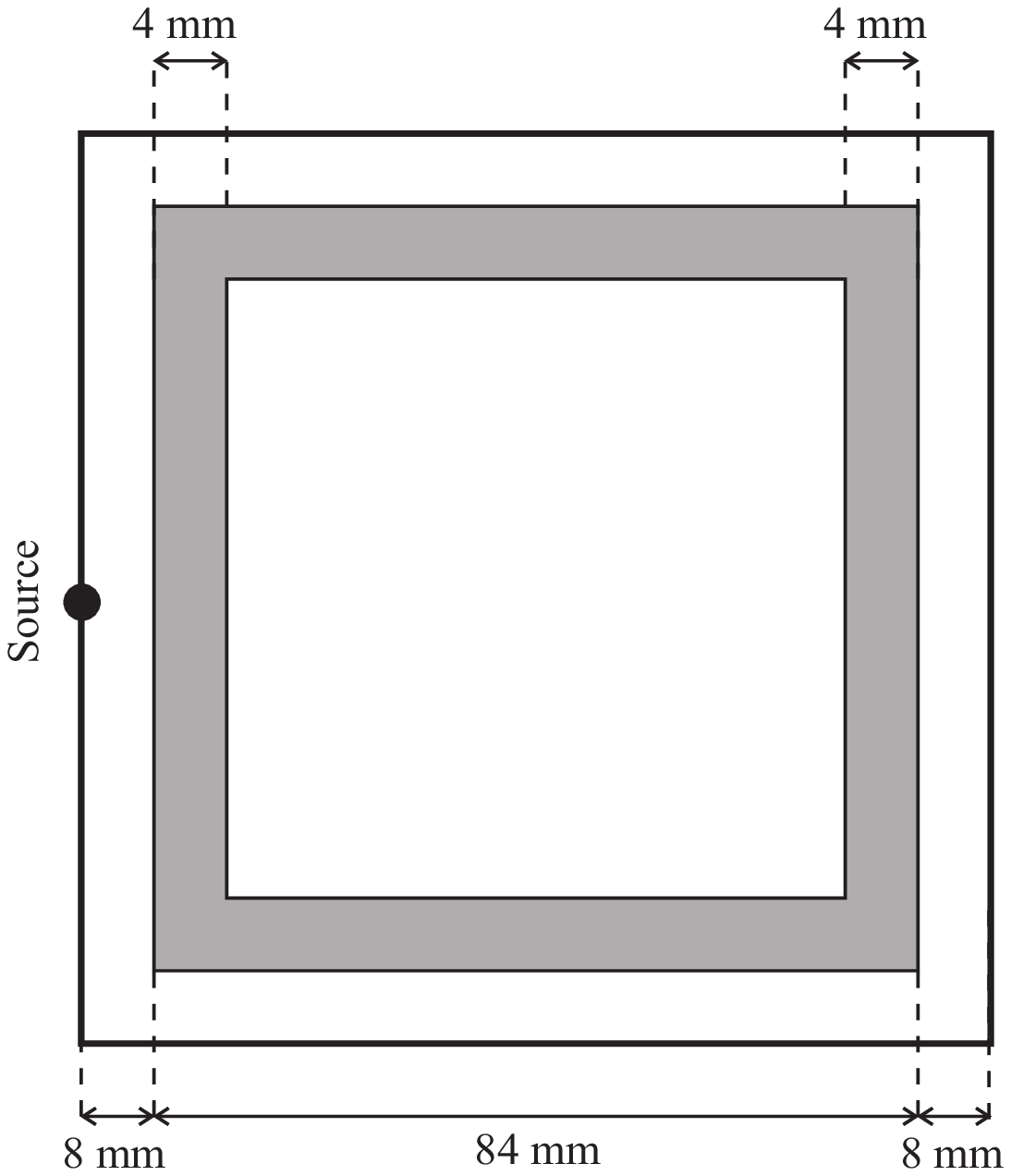}\label{fig:RechteckAydin}}
 \hfill \subfigure[Cut along $y=50$ mm.]{
  \includegraphics[width=0.53\linewidth]{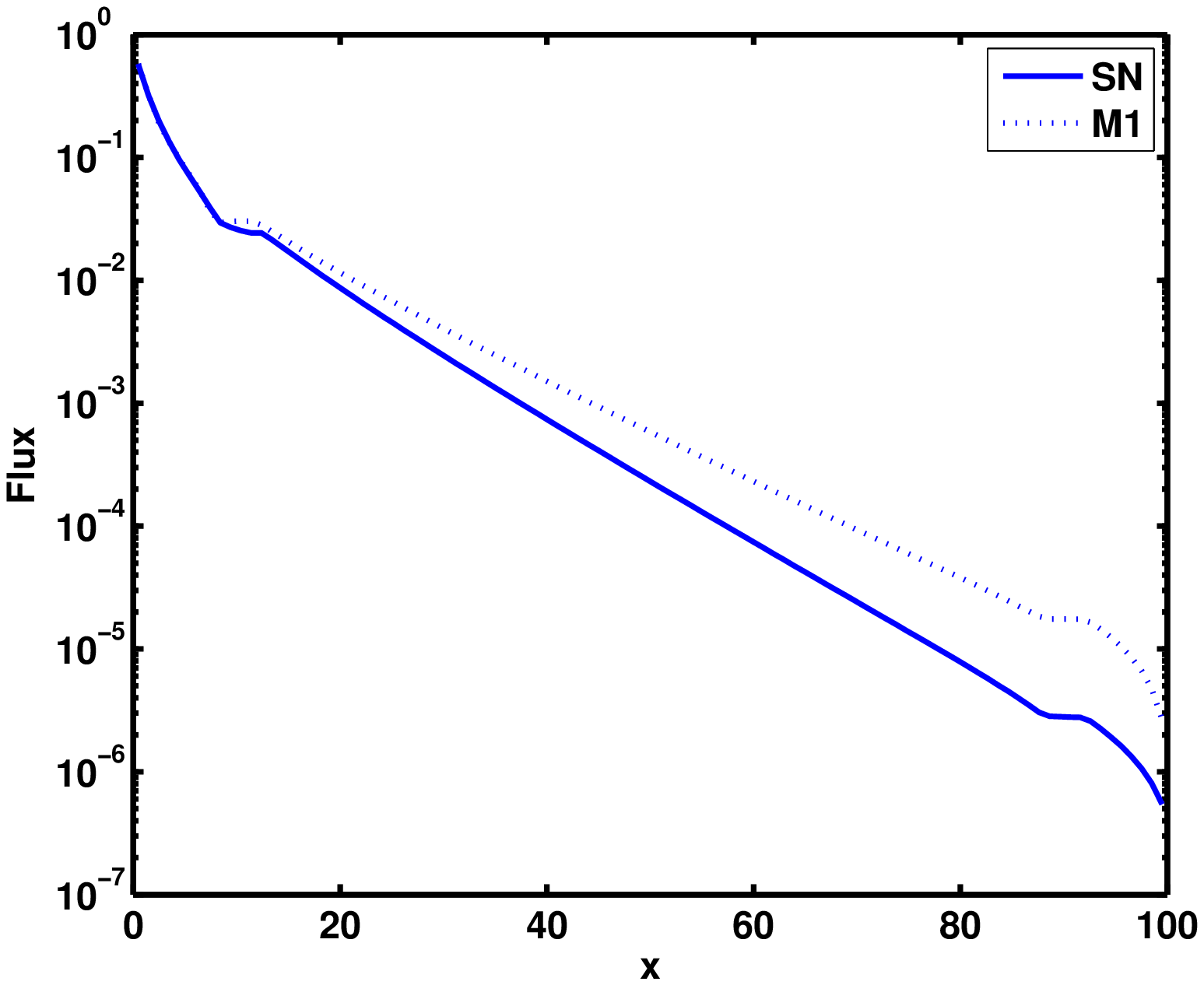}\label{fig:Cut}}\\
  \subfigure[Transport solution.]{
  \includegraphics[width=0.49\linewidth]{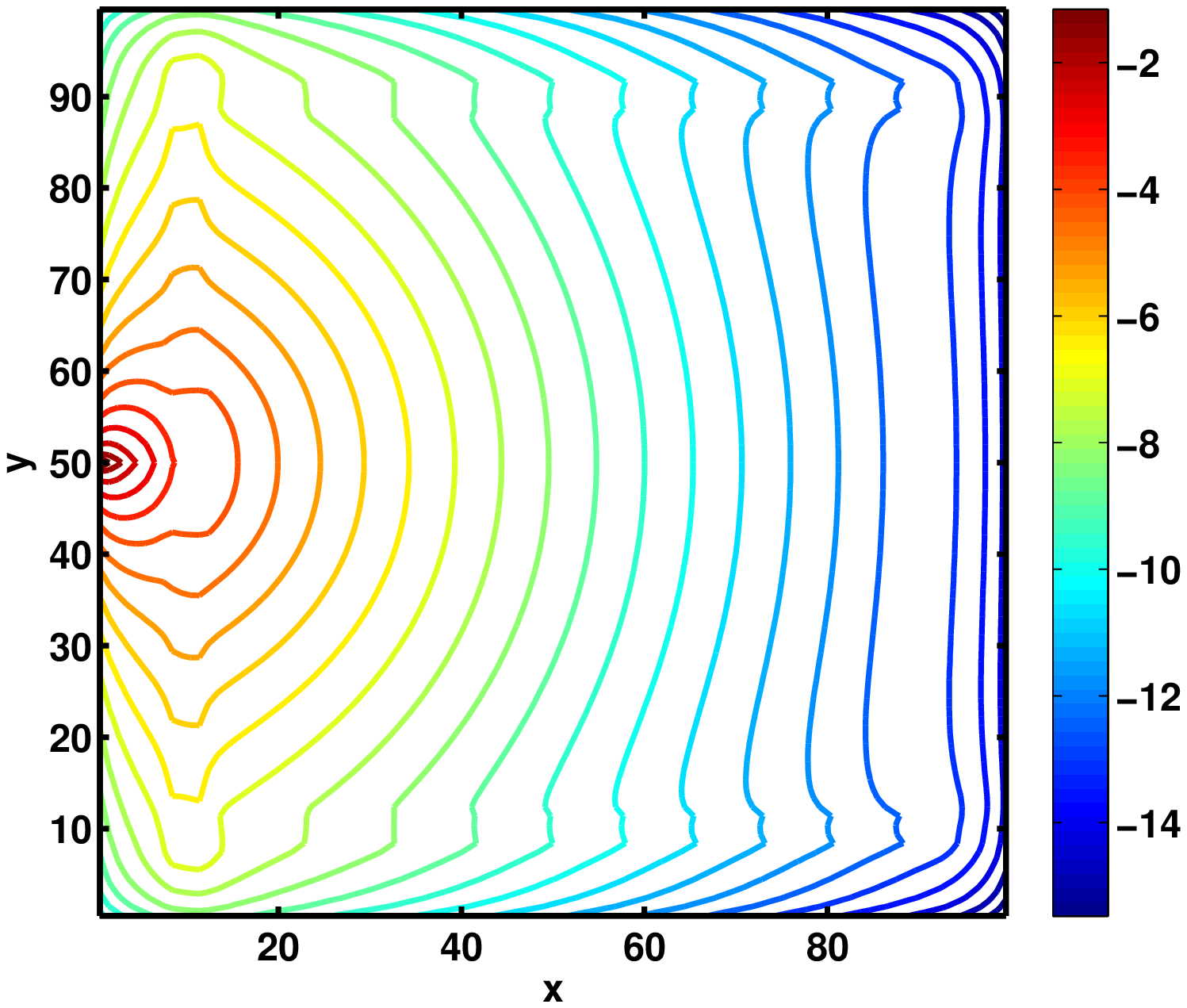}}
  \subfigure[Minimum entropy model.]{
\hfill  \includegraphics[width=0.49\linewidth]{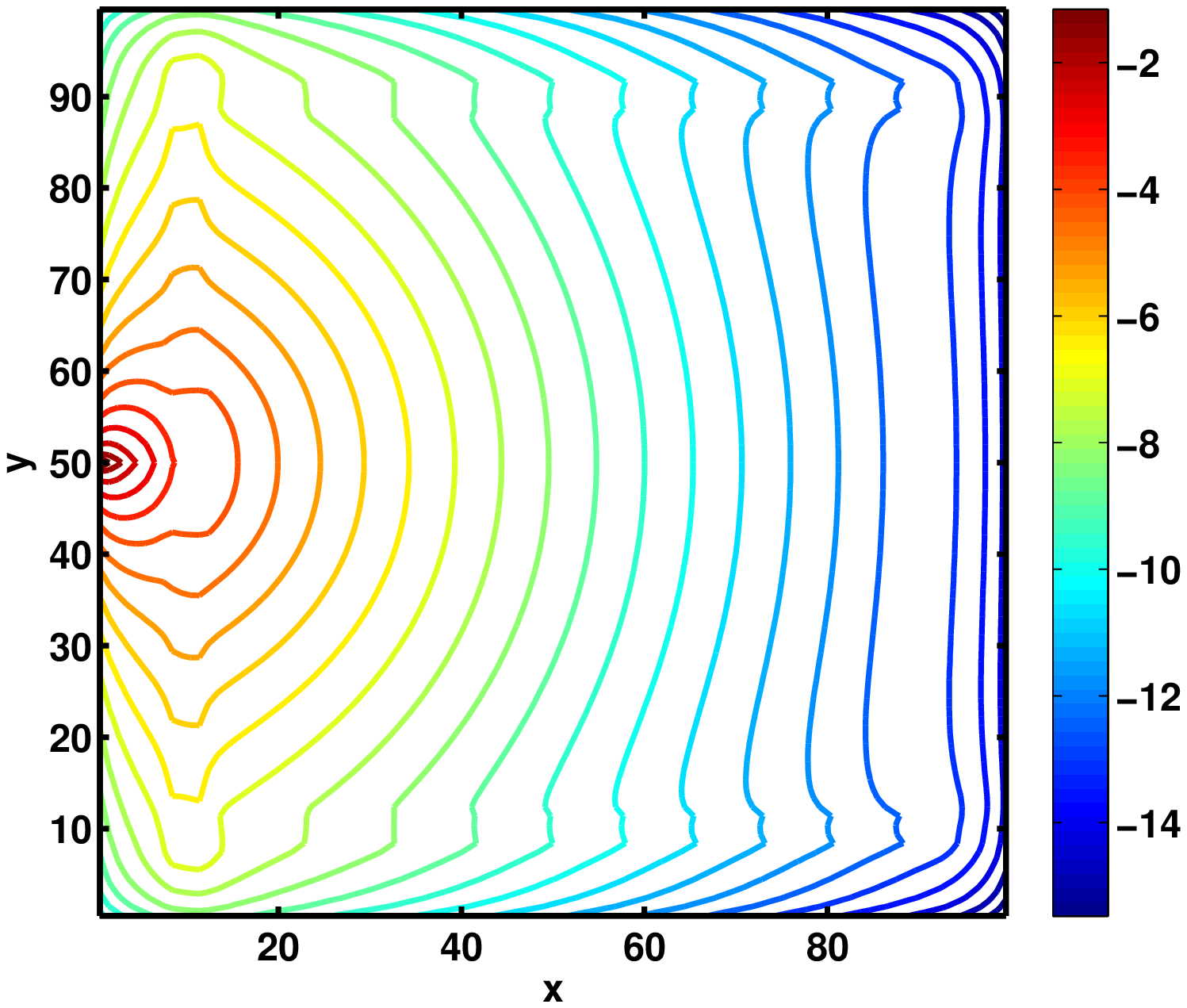}}
    \caption{Transport versus minimum entropy for void-like layer.}
  \label{fig:AydFig12}
\end{figure}

Our second test case, again taken from \cite{AydOliGod02}, is a two-dimensional quadratic domain which contains a void-like layer, shown in gray in Figure \ref{fig:RechteckAydin}. 
Again, we consider only elastic scattering modeled by the Henyey-Greenstein kernel.

We take $\sigma_s=0.5$ mm$^{-1}$ and $\sigma_a=0.005$ mm$^{-1}$ inside
the square, and $\sigma_s=0.01$ mm$^{-1}$ and $\sigma_a=0.0001$
mm$^{-1}$ in the void-like ring. In both regions, $g=0$. An isotropic
source of particles is placed on the left boundary.  In a 2D contour
plot (Figure \ref{fig:AydFig12}), the fluxes $\psi^{0}$ from the
discrete ordinate method and from the minimum entropy method are
virtually indistinguishable. The propagation into the medium, as well
as the void-like layer are equally well resolved. A difference between
the models only becomes apparent in a logarithmic plot of a cut
through the center of the square at $y=50$ mm. Figure
\ref{fig:AydFig12} shows the particle flux along this line. The
difference between both solutions is again of the order of one
percent.

\subsection{Electrons on Water Phantom}
\begin{figure}
\centering\includegraphics[width=0.7\linewidth]{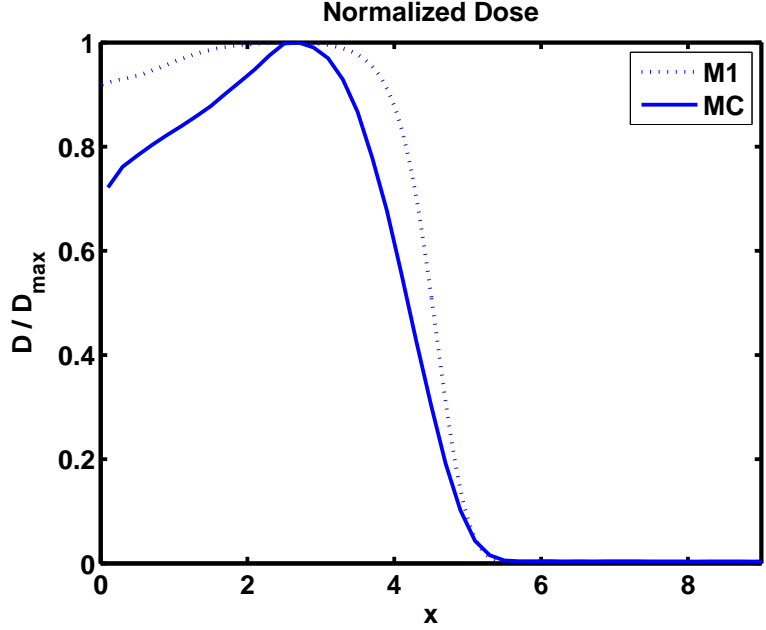}
\caption{Dose for 10 MeV electron beam on water.}
\label{fig:M1MC}
\end{figure}

As a first test case that includes energy loss, we consider a 10 MeV
electron beam impinging onto a slab of water. In Figure \ref{fig:M1MC}
we compare the results computed with our code to the dose computed by
the state-of-the-art Monte Carlo code PENELOPE
\cite{SalFerSem08}. This code has been extensively validated against
experimental results.

To obtain a good fit with the tabulated scattering data, we have fixed
our model parameters for water as $\epsilon_B = 16.0$ eV, $Z=9.40$,
$\rho_{\rm el}=0.256\times 10^{23}$ g/cm$^3$, $\rho_{\rm in} =
6.21\times 10^{23}$ g/cm$^3$. These parameters are directly inserted
into the model \eqref{eq:BCSD}, and subsequent derived models issued
from \eqref{eq:BCSD}. As boundary conditions, we have taken a very
narrow Gaussian in energy, and a $\delta$ pulse in angle
$$
\psi_b = \psi_0 \exp(-200(\epse-\epse_{\rm beam})^2) \delta(\mu-1),
$$ and computed the angular moments. PENELOPE was set up in a
pseudo-1D setting with a large beam size perpendicular to the beam
direction.

In order to compare the different formulations of the models, both
depth-dose curves in Figure \ref{fig:M1MC} have been normalized to
dose maximum one. The penetration depth computed with the $M_1$ model
agrees very well with the Monte Carlo result. In fact this deviation
is within the margin of differences between different Monte Carlo
codes \cite{SemWilBie00}. The only major difference occurs near the
boundary, where the $M_1$ model overestimates the dose. This might be
due to the simplified physics (possibly neglection of Bremsstrahlung
effects) or an oversimplification of the angular dependence of $\psi$
in the $M_1$ model. Both possibilities will be investigated
further. However, we believe that this result can serve as a proof of
concept of a PDE based modeling of dose computation.

\subsection{CT Data}
\begin{figure}
\centering\includegraphics[width=0.5\linewidth]{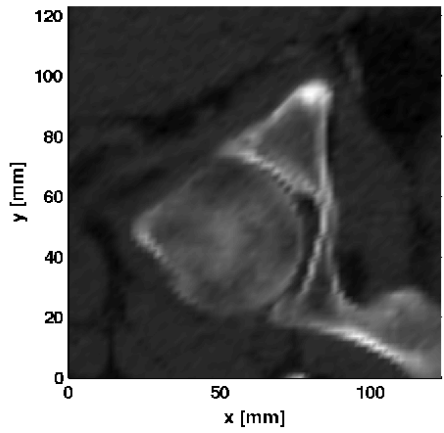}
\caption{CT data of hip bone.}
\label{fig:CT}
\end{figure}

\begin{figure}
\subfigure[Monte Carlo solution.]{\includegraphics[width=0.48\linewidth]{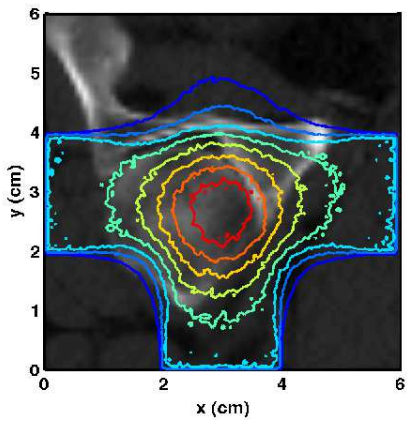}}
\subfigure[Minimum entropy solution.]{\includegraphics[width=0.48\linewidth]{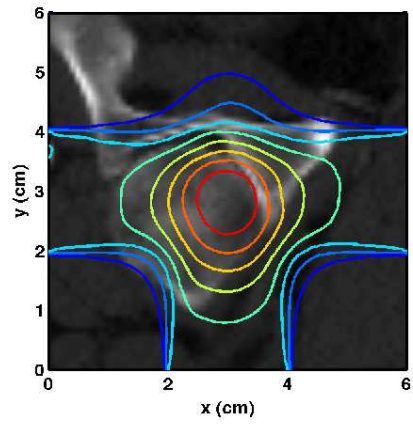}}
\subfigure[Cut along $x=3$ cm.]{\includegraphics[width=0.48\linewidth]{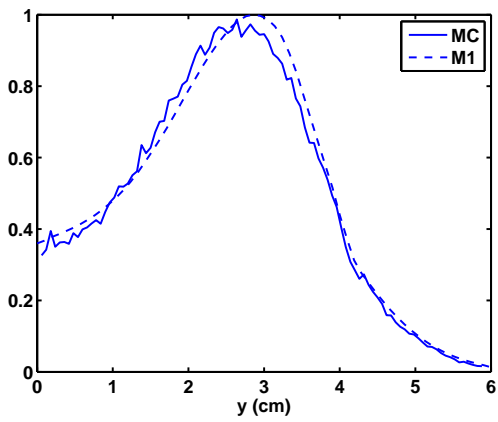}}
\subfigure[Cut along $y=3$ cm.]{\includegraphics[width=0.48\linewidth]{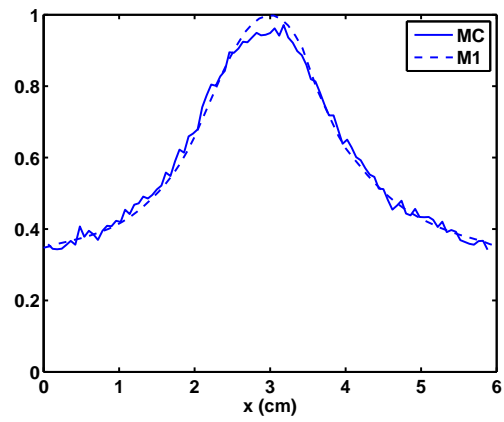}}
\caption{Dose distribution for three beams impinging on hip bone.}
\label{fig:Dose_CT}
\end{figure}

In our final test case, we compare our method with Monte Carlo results
from PENELOPE using real patient CT data showing the hip bone.  We
took a two-dimensional slice of $6\times 6$ cm from the
three-dimensional CT data. A square region is split into 64$\times$64
squares. In each of the squares, the material is described by its
Hounsfield grey value $ \mathcal{G}(x,y)$. The grey values can be
translated into physical parameters as follows,
$$
\rho(x,y) = \left(\frac{\mathcal{G}(x,y)}{1000}+1\right) \rho_\text{Water},
$$
{\it i.e.}\ the densities $\rho_{\rm el}$ and $\rho_{\rm in}$ for water are multiplied by a specified factor. The region shows the hip bone and the density varies between 86\% and 226\% of the value of water.
The boundary conditions were set up similar to the previous case, with three beams of width 2 cm, each consisting of 10 MeV electrons, impinging from the centers of three sides of the domain. Contour plots of the dose distribution are shown in Figure \ref{fig:Dose_CT}. There, we also show two cuts through the dose distribution. looking at the 2D dose distribution, the contour lines agree very well. Note that, although we have used $3\times 10^{10}$ particles, there still is significant noise in the Monte Carlo results. The two cuts through the beam centers show that also quantitatively the independently computed dose distributions agree very well.

The computation time for the 3D Monte Carlo dose was $3\times 29$ hours for $3\times 10^{10}$ particles on a 3GHz Pentium 4 with 1 GB RAM. In 1D, the minimum entropy model took about 1 second, in 2D 4 seconds. Thus we expect a computation time of several seconds in a full 3D dose computation. 

Again, this result shows that if our model is developed further, it may serve as an alternative to existing dose computation methods.

\section*{Acknowledgements}
We thank Edgar Olbrant for performing the PENELOPE computations. The CT data has been provided by the German Cancer Research Center DKFZ and the Optimization Department at Fraunhofer ITWM, Kaiserslautern. This work was supported by the German Research Foundation DFG under grant KL 1105/14/2, by the French Ministry of Foreign Affairs under 
EGIDE contract 17852SD and by German Academic Exchange Service DAAD under grant D/0707534.



%





%

\bibliographystyle{amsplain}
\bibliography{RadLit}

\end{document}